\documentstyle[12pt,epsf,epsfig,wrapfig]{article}
 \newcommand{\be}{\begin{eqnarray}}
 \newcommand{\ee}{\end{eqnarray}}
 \newcommand{\nn}{\nonumber\\}
\begin{document}

\begin{center}
{\bf NON-SINGLETS IN
 \\
SEMI-INCLUSIVE DIS AND INCLUSIVE $E^+E^-$ ANNIHILATION
}
\footnote{This work is supported by a Collaborative Grant of the UK Royal Society}
\\
\vspace{.5cm}

{
\underline{Ekaterina Christova}$^\dagger$,
  Elliot Leader$^{\dagger\dagger}$}\\
{\it $\dagger$ Institute for Nuclear Research and Nuclear Energy,
Sofia,  echristo@inrne.bas.bg\\
$\dagger\dagger$ Imperial College,  London,  e.leader@imperial.ac.uk}
\end{center}
\vspace{0.1cm}

\begin{abstract}
We show that non-singlets in semi-inclusive DIS  with $\pi^\pm$ determine without assumptions
$\Delta u_V$, $\Delta d_V$ and $D_u^{\pi^+-\pi^-}$. Non-singlets in SIDIS
 and inclusive $e^+e^-$-annihilation with $K^\pm$ determine $s-\bar s$ and $\Delta s-\Delta \bar s$, but an assumption is
 necessary -- we choose  as most natural $D_d^{K^+-K^-}=0$.
 Measurements with charged $K^\pm$ and neutral $K^0+\bar K^0$ in semi-inclusive DIS  and $e^+e^-$-annihilation  determine
 $D_{u,d,s}^{K^++K^-}$ without any assumptions and allow tests of LO approximation in QCD.
\end{abstract}
\vspace{.8cm}


{\bf 1.} The difference between the spin-dependent structure functions on
proton and neutron, $g_1^p$ and $g_1^n$,  measures the
non-singlet (NS) flavour combination $\Delta q_3= (\Delta
u+\Delta\bar u) - (\Delta u+\Delta\bar u)$:
\be
g_1^p-g_1^n=
\frac{1}{6}\Delta q_3\left( 1+\frac{\alpha_s}{2\pi} +...\right).
\ee
This measurable quantity is free of any uncertainties and,
being a NS, this holds in all orders of QCD. Actually, $\Delta
q_3$ is the only quantity determined in inclusive DIS without any
assumptions.

In general, for all quantities that measure NS combinations we
have that first, in all orders in QCD and second, in their
$Q^2$-evolution, no new parton densities appear. That last one is
especially useful as it allows to compare directly measurable
quantities at different $Q^2$ and different experiments.

We consider measurable quantities that single out NS's in
semi-inclusive DIS (SIDIS), both polarized and unpolarized, and
in inclusive $e^+e^-$ annihilation:
\be
e+N \to e+h+X,\qquad e^++e^-\to h+X
\ee
and discuss the non-perturbative
information  about the parton distribution (pdf) and fragmentation (ff) functions one can obtain.
\vspace{0.5cm}

{\bf 2.}
Recently we pointed out~\cite{SPIN2004} that the difference cross section in SIDIS
measures a NS both in pdf's and in ff's. This follows immediately from $C$-inv. of the ff's:
\be
D_G^{h-\bar h}=0,\quad D_q^{h-\bar h}= -D_{\bar q}^{h-\bar h},\quad
D_f^{h-\bar h}\equiv D_f^h - D_f^{\bar h}.
\ee
Here $\bar h$ is the $C$-conjugate of the hadron $h$. Then for the difference cross section
$\sigma_N^{h-\bar h}\equiv
\sigma_n^h-\sigma_N^{\bar h}$, we obtain that in all orders in QCD gluons cancel, i.e. no $g(x)$ and no $D_g^h(z)$:
\be
 \tilde \sigma_N^{h-\bar h}(x,z) =\frac{1}{9}\left[4  u_V\otimes D_u^{h-\bar h}
+  d_V\otimes D_d^{h-\bar h}
+ (s -\bar s)\otimes D_s^{h-\bar h}\right]\otimes  \hat\sigma_{qq} (\gamma
q \to q X).\label{diff}
\ee
Here $\hat\sigma_{qq}$ is  the perturbatively  QCD-calculable partonic cross section
$q\gamma^*\to q+X$:
\be
 \hat\sigma_{qq} &=&  \hat\sigma_{qq}^{(0)} + \frac{\alpha_s}{2\pi} \hat\sigma_{qq}^{(1)} +...
\ee
Thus,  $ \tilde \sigma_N^{h-\bar h}$ is a NS in both the pdf's and ff's,
and in addition -- each term is such a NS. Eq.(\ref{diff}) shows that it
is sensitive only to the valence quark pdf's
and $(s-\bar s)$. Depending on the properties of the final measured hadrons
$h-\bar h$, one can single out different pieces of $ \tilde \sigma_N^{h-\bar h}$.
\vspace{0.5cm}

{\bf 3.}
It is clear that the above arguments hold for polarized SIDIS as well, except that the unpolarized
pdf's  and unpolarized partonic cross sections are replaced by the polarized ones.
Then,  measuring the difference polarization asymmetry $A_{N}^{h-\bar h}$~\cite{SPIN2004}:
\be
A_{N}^{h-\bar h}(x,z) = \frac{1+(1-y^2)}{2y\,(2-y)}\frac{\Delta\sigma_N^{h}-\Delta\sigma_N^{\bar h}}
{\sigma_N^{h}-\sigma_N^{\bar h}}
\ee
one obtains information about $\Delta u_V$, $\Delta d_V$ and $(\Delta s-\Delta\bar s)$
 without requiring any knowledge about $\Delta g$, $g$ and $D_g^h$. If $h=\pi^\pm$ one determines
  $\Delta u_V$ and  $\Delta d_V$ without knowledge even about the stange quarks $s$ and $\Delta s$
  ( due to SU(2) symmetry of the pions $D_s^{\pi^+-\pi^-}=0)$. In JLab this asymmetry will be
  measured with enough accuracy~\cite{JLab}.
  If $h=K^\pm$ measuring $A_{N}^{K^+-K^-}$
  one can obtain information
  about $(\Delta s -\Delta\bar s)$, provided we have determined $\Delta u_V$ and  $\Delta d_V$ in
  $A_{N}^{\pi^+-\pi^-}$.

  In (\ref{diff}) the ff's $D_q^{h-\bar h}$ enter. Usually these quantities are simulated by
  the LUND model. As pointed out in \cite{Stefan},
  and very recently also in \cite{deFlorian}, the result of the analysis for the polarized pdf's
  (especially for the sea-quarks) depend on the precision of our knowledge of the ff's.
  Our goal is to discuss  the information that one can obtain about ff's
  using only measurable quantities from unpolarized SIDIS and/or inclusive $e^+e^-$-annihilation.
  We shall show that when analyzing $A_{N}^{K^+-K^-}$ an  additional assumption is needed. We
  choose as most reasonable one $D_d^{K^+} = D_d^{K^-}$.
\vspace{0.5cm}

{\bf 4.}
Following (\ref{diff}), measurements of the difference ratio $R_N^{h -\bar h}$ of unpolarized SIDIS to inclusive DIS:
\be
R_N^{h -\bar h}=\frac{\sigma_N^{h}-\sigma_N^{\bar h}}{\sigma_N^{DIS}}
\ee
can be used to determine, in any order in QCD,  the difference of the ff's, $D_q^{h-\bar h}$,
and to obtain information about $s-\bar s$.

If $h=\pi^\pm$,  SIDIS on protons and neutrons provide  two independent measurements to determine
 $D_u^{\pi^+-\pi^-}$  (SU(2) inv. implies $D_u^{\pi^+-\pi^-}$= -$D_d^{\pi^+-\pi^-}$).

It is more complicated if $h=K^\pm$, because we cannot use SU(2) inv. to reduce the number
of ff's. If we assume $D_d^{K^+-K^-}=0$, then measurements on $p$ and $n$ provide two measurements for
$D_u^{K^+-K^-}$ and $(s-\bar s)\otimes D_s^{K^+-K^-}$. Note that because $D_s^{K^+-K^-}$ is presumably
 not a small quantity ($s$ is a valence quark for $K^-$ and non-valence quark for $K^+$),
 the product  $(s-\bar s)\otimes D_s^{K^+-K^-}$ can be zero
(non-zero) only if $s-\bar s =0$ ($s-\bar s \neq 0$).

>From the quark content of $K^\pm$, the assumption $D_d^{K^+-K^-}=0$ seems very reasonable
if  $K^\pm$ are directly produced. However, if resonances have to be taken into account, and $K^\pm$
 appear also as decay products of the resonances (which
is very often the real situation in experiment) this assumption is not undisputed. It's a bad luck that,
 as we shall show, even inclusive $e^+e^-\to K^\pm X$ does not help to determine $D_d^{K^+-K^-}$.
 However it helps to determine other NS's.
\vspace{0.5cm}

{\bf 5.}
 In general, the cross section for inclusive $e^+e^- \to hX$  is~\cite{Rijken}:
\be
\frac{d\sigma^h}{dz\,d\cos\theta} = \frac{3}{8}(1+\cos^2\theta )\,d\sigma_T^h(z) +
\frac{3}{4}(1-\cos^2\theta )\,d\sigma_L^h(z)
+ \frac{3}{4}\cos\theta \,d\sigma_A^h(z)
\ee
where $d\sigma_{T,L,A}^h$ are the transverse, longitudinal and asymmetric cross sections:
$d\sigma_{T,L,}^h$ are measured through the total (integrated over $\cos\theta$) cross section, $d\sigma_{A}^h$
is singled out through the forward-backward asymmetry. In LO we have ($d\sigma_L^h$ does not contribute in LO):
\be
d\sigma^h_T(z)&=&\int_{-1}^1 d\cos \vartheta \left(\frac{d\sigma^h}{dz\,d\cos\theta}\right) =
3\,\sigma_0\,\sum_q \hat e_q^2\,D_q^{h+\bar h},\quad \sigma_0=\frac{4\pi\alpha_{em}^2}{3\,s}\\
A_{FB}^h(z)&=&\left[\int_{-1}^0 -\int_0^1 \right] (d\cos
\vartheta \left(\frac{d\sigma^h}{dz\,d\cos\theta}\right)
=3\,\sigma_0\,\sum_q \frac{3}{2}\,\,\hat a_q\,D_q^{h-\bar h} \ee
i.e. $d\sigma^h_T(z)$ measures $D_q^{h+\bar h}$, while 
$A_{FB}^h(z)$ measures the NS's $D_q^{h-\bar h}$ we are
interested in. Assuming that the photon  and the neutral
$Z^0$-boson are exchanged we have: \be \hat{e_q}^2(s) &=&e_q^2 -
2e_q \,v_e\,v_q\,\Re e \,h_Z + (v_e^2 + a_e^2) \, \left[(v_q)^2
+(a_q)^2\right]\, \vert h_Z\vert ^2\nn \hat a_q &=&
2\,a_e\,a_q\,\left(-e_q\, \Re e\,h_Z + 2\,v_e\,v_q\,\vert
h_Z\vert^2\right). \ee Here $e_q$ is the electric charge of the
quark $q$ and \be
 v_e&=&-1/2 +2 \sin^2\theta_W,\quad a_e=-1/2, \nn
v_q&=&I_3^q-2e_q\sin^2\theta_W,\quad a_q=I_3^q, \quad I_3^u = 1/2, \quad I_3^d = -1/2.\nn
h_Z &=& [s/(s-m_Z^2+im_Z\Gamma_Z)]/\sin ^2 2\theta_W.
\ee

If $h=K^\pm$ we have
\be
\frac{A_{FB}^{K^+-K^-}}{3\,\sigma_0} =
\frac{3}{2}\, \left[\hat a_u\,D_u^{K^+-K^-} + \hat a_d\,(D_d^{K^+-K^-} + D_s^{K^+-K^-})\right].\label{AFBK}
\ee

If we combine this measurement with the SIDIS measurements $R_N^{K^+-K^-}$ on unpolarized $p$ and $n$
 (see (eq.(\ref{diff}), $h=K^\pm$)
we obtain 3 measurements for the 4 unknown quantities $D_{u,d,s}^{K^+-K^-}$ and ($s-\bar s$).
We need an assumption: either
 $s=\bar s$ or $D_d^{K^+-K^-}=0$. We consider  $D_d^{K^+-K^-}=0$ as a reasonable one.
Note that up to now all analysis of the experimental data have
been performed assuming $s=\bar s$ and measurements of
$R_N^{K^+-K^-}$ will offer a real possibility to justify this
assumption.
 \vspace{0.5cm}

{\bf 6.} If in unpolarized SIDIS and $e^+e^-$-inclusive processes,
in addition to the charged $K^\pm$
   also the neutral $K^0 +\bar K^0$ are  observed,
  we can use SU(2) invariance to relate the neutral to the charged Kaon ff's:
\be
D_u^{K^+ + K^-}=D_d^{K^0+ \bar K^0},\qquad
D_d^{K^+ + K^-}=D_u^{K^0+ \bar K^0},\quad
D_s^{K^+ + K^-}=D_s^{K^0+ \bar K^0}.
\ee
and no new ff's appear in the cross sections. In LO we have:
\be
d\sigma^{K^+ +K^-- (K^0 +\bar K^0)}_T(z)&=&
3\,\sigma_0\left[(\hat e^2_u - \hat e^2_d )_{m_Z^2} (D_u - D_d)^{K^+ +K^-}\right]\label{a}\\
d\tilde\sigma_p^{K^+ +K^-- (K^0 +\bar K^0)}(x,z) &=&\frac{1}{9} \,[4(u+\bar u) -(d+\bar d)] (D_u - D_d)^{K^+ +K^-}\label{b}\\
d\tilde\sigma_n^{K^+ +K^-- (K^0 +\bar K^0)}(x,z) &=&\frac{1}{9}\, [4(d+\bar d) -(u+\bar u)](D_u - D_d)^{K^+ +K^-}\label{c}
\ee
i.e. we have three measurements -- $R_{p,n}^{K^+ +K^-- (K^0 +\bar K^0)}$ and $d\sigma^{K^+ +K^-- (K^0 +\bar K^0)}_T$
 -- to determine  $(D_u - D_d)^{K^+ +K^-}$,
 but presumably at different $Q^2$. As $(D_u - D_d)^{K^+ +K^-}$ is a NS we can easily evolve it in $Q^2$ and
 form possible tests of the LO approximation. For example:
 \be
\frac{9\,d\tilde{\sigma}_p^{K^++K^--(K^0+\bar K^0)}(x,z,Q^2)}
{d\sigma_T^{K^++K^--(K^0+\bar K^0)}(z,m_Z^2)_{ \downarrow Q^2}}=
\frac{[4(u+\bar u) -(d+\bar d)](x,Q^2)}{3\,\sigma_0\,(\hat e_u^2 - \hat e_d^2)_{m_Z^2}}
\ee
Here $d\sigma_T^{K^++K^--(K^0+\bar K^0)}(z,m_Z^2)_{ \downarrow Q^2}$ denotes that the data is
measured at $m_Z^2$ and then
evolved to $Q^2$ according to the DGLAP equations. As this is a NS, the evolution
does not introduce any new quantities. Combined with measurements of any of the two
quantities $d\sigma^{K^+ +K^-+ (K^0 +\bar K^0)}_T(z)$ or
$R_{p,n}^{K^+ +K^-+ (K^0 +\bar K^0)}(x,z)$, we obtain enough measurements to determine $D_{u,d,s}^{K^+ +K^-}$
without any assumptions (recall that ($s+\bar s$) is known from DIS measurements).
Contrary to $d\sigma^{K^+ +K^- -(K^0 +\bar K^0)}_T(z)$,
though the total cross section
 $d\sigma^{K^+ +K^-+ (K^0 +\bar K^0)}_T(z)$ is very precisely measured at $s=m_Z^2$,  its
 $Q^2$-evolution involves the poorly known $D_g^K$  and large errors are introduced when combined with the
 SIDIS measurements performed
   at much lower $Q^2$~\cite{Stefan}.
Note that  when forming the cross sections for the difference $K^++K^--(K^0+\bar K^0)$,
measurement of  the neutral Kaons is essential in order to cancel the $s$-quark contributions,
while for the sum $K^++K^-+(K^0+\bar K^0)$
the neutral Kaons will only improve the statistics.

\end{document}